\documentclass[aps,prd,floatfix,nofootinbib,superscriptaddress,showpacs,twocolumn]{revtex4-1}

\usepackage{amssymb}
\usepackage[intlimits]{amsmath}
\usepackage{amsfonts}
\usepackage{dsfont}
\usepackage{subfigure}
\usepackage[colorlinks]{hyperref}
\usepackage{graphicx}
\usepackage[usenames]{color}
\usepackage{natbib}
\usepackage{multirow}
\usepackage{verbatim}

\newcommand{\sh}[1]{#1\hskip-7pt \diagup}
\newcommand{\Sh}[1]{#1\hskip-10pt \diagup}

\def\i{\ensuremath{\mathrm{i}}}

\def\T{\ensuremath{\mathrm{T}}}

\newcommand{\QCD}{\ensuremath{\mathrm{QCD}}}

\newcommand{\Eq}[1]{Eq.~(\ref{#1})}

\begin{document}

\title{The role of momentum dependent dressing functions and vector meson dominance
in hadronic light-by-light contributions to the muon g-2. }

\author{Tobias Goecke}
\affiliation{Institut f\"ur Theoretische Physik, 
Universit\"at Giessen, 35392 Giessen, Germany}
\author{Christian S. Fischer}
\affiliation{Institut f\"ur Theoretische Physik, 
 Universit\"at Giessen, 35392 Giessen, Germany}
\affiliation{Gesellschaft f\"ur Schwerionenforschung mbH, 
  Planckstr. 1  D-64291 Darmstadt, Germany.}
\author{Richard Williams}
\affiliation{Institut f\"ur Physik, Karl-Franzens--Universit\"at Graz, Universit\"atsplatz 5, 8010 Graz, Austria.}

\begin{abstract}

  We present a refined calculation of the quark-loop contribution to hadronic 
  light-by-light scattering that focuses upon the impact of the transverse
  components of the quark-photon vertex. These structures are compared and 
  contrasted with those found within the extended NJL-models. We discuss
  similarities and differences between the two approaches and further clarify 
  the important role of momentum dependent dressing functions.   
  As expected we find that the transverse structures of the quark-photon
  vertex lead to a suppression of the quark-loop contribution to the anomalous 
  magnetic moment of the muon. However, we find evidence that this suppression 
  is overestimated within models with simple approximations for the quark-photon 
  interaction.  
\end{abstract}

\maketitle


%
%
\section{Introduction}
\begin{figure}[t]
  \begin{center}
    \includegraphics[width=0.15\textwidth]{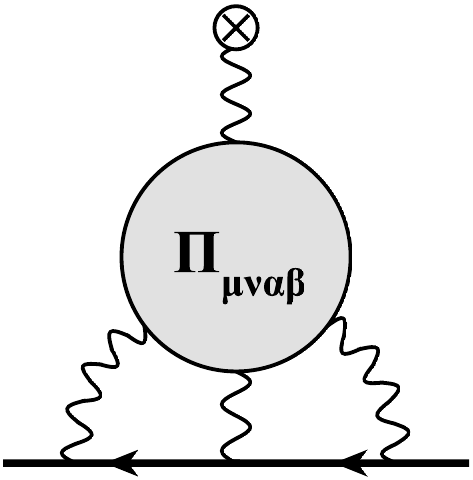}
  \end{center}
  \caption{The light-by-light scattering contribution to the muon $g-2$.
  The main ingredient is the hadronic photon four-point function $\Pi_{\mu\nu\alpha\beta}$ to be
  discussed below.}
  \label{fig:LBLContribution}
\end{figure}

\begin{table}[b]
  \centering

  \begin{tabular}{c||r|l|l}
    Contribution & $\hspace{-0.6cm}a_\mu\times 10^{11}\hspace{0.5cm}$ & 
    $\hspace{0.5cm}\dfrac{a_\mu^i}{a_\mu^{SM}}$     & $\left(\dfrac{\delta a_\mu^i}{\delta a_\mu^{SM}}\right)^2$  \\
    \hline\hline
    QED		&	$116\,584\,718.1\,(\,\,\,0.2)$ & $99.99390\%$ & $00.00098\%$   \\
    \hline
    weak	&	$153.2\,(\,\,\,1.8)$ & $00.00013\%$ & $00.07910\%$ \\
    \hline
    QCD LOHVP	&	$6\,949.1\,(58.2)$ & $00.00596\%$ & $82.69628\%$ \\
    \hline
    QCD HOHVP	& 	$-98.4\,(\,\,\,1.0)$ & $00.00008\%$ & $00.02441\%$	\\
    \hline
    QCD LBL	&	$105\,\,\,\,\,\,(26\,\,\,\,)$ & $00.00009\%$ & $16.50391\%$ \\
    \hline
    Standard Model &   $116\,591\,827.0\,(64\,\,\,\,)$ & $100\%$ & $100\%$ \\
    \hline
    Experiment	&	$116\,592\,089\,\,\,\,\,\,(63\,\,\,\,)$ &\\
    \hline\hline
    Exp-Theo	&	$262\,\,\,\,\,\,(89\,\,\,\,)$ &
  \end{tabular}
    \caption{Standard Model contributions to the muon $g-2$.}
  \label{tab:DiffContrToAm}
\end{table}
The anomalous magnetic moment of the muon $a_\mu$ is an observable that furnishes
a precision test of the electromagnetic (EM) interaction, the weak interaction 
and the strong interaction. The relative size of the different theoretical parts 
and comparison with the experimental results of the E821 experiment at 
Brookhaven~\cite{Bennett:2006fi,Roberts:2010cj} are shown in 
table~\ref{tab:DiffContrToAm}. The dominant contribution of more than $99\%$ is 
due to the EM interaction as described by quantum electrodynamics (QED), which 
has been evaluated up to order $\alpha^5$ in the fine structure constant 
\cite{Aoyama:2012wk}. In addition, given the theoretical and experimental 
precision available, the weak interaction yields significant contributions 
\cite{Czarnecki:2002nt}. 
An interesting feature of the observable $a_\mu$ is its sensitivity to non-perturbative 
QCD corrections. The leading QCD contribution amounts to about $0.006\%$ which is about 
two orders of magnitude larger than the total theory uncertainty. Together, the leading 
and sub-leading QCD contributions dominate the error of the 
total Standard Model prediction as can be seen in the table. This is because 
perturbation theory is not applicable and the methods used lead to a 
substantial error here. The two most relevant hadronic contributions are the hadronic 
vacuum polarisation (HVP) and the light-by-light (LBL) scattering contribution. The 
former is related to experimentally available $e^+ e^- \rightarrow hadrons$ data via 
dispersion relations (see \cite{Jegerlehner:2009ry}) and as such the error can be 
systematically reduced. This quantity is furthermore the subject of several recent 
lattice studies~\cite{Feng:2011zk,Boyle:2011hu,DellaMorte:2011aa}. Furthermore, we 
have applied the method of Dyson-Schwinger equations (DSEs) to this quantity and were 
able to reproduce the dispersion analysis result on the ten percent 
level~\cite{Goecke:2011pe}. The result for HVP quoted in Table~\ref{tab:DiffContrToAm} 
is taken from Ref.~\cite{Hagiwara:2011af}.
Whilst presently the uncertainty is dominated by HVP, the LBL contributions 
(shown in Fig.~\ref{fig:LBLContribution}) are potentially more problematic in the long 
run since it is extremely hard to determine these in a model-independent way. The LBL
contributions have been investigated from the view point of low-energy effective 
models such as the 
Extended Nambu--Jona-Lasinio (ENJL) model~\cite{Bijnens:1995xf}, the Hidden Local 
Symmetry (HLS) model~\cite{Hayakawa:1995ps}, vector meson dominance (VMD) 
approaches~\cite{Knecht:2001qf,Melnikov:2003xd}, the non-local chiral quark 
model (NL$\chi$QM) \cite{Dorokhov:2008pw,Dorokhov:2012qa}, the chiral constituent 
quark model ($\chi$CQM)~\cite{Greynat:2012ww}, in holographic models~\cite{Cappiello:2010uy} 
and Dyson-Schwinger Equations~\cite{Fischer:2010iz,Goecke:2010if}. The lattice 
calculations of LBL are still at an exploratory stage \cite{Hayakawa:2005eq}. 
The LBL-contribution quoted in Table~\ref{tab:DiffContrToAm} is taken from Ref.
\cite{Prades:2009tw}. There different groups, pursuing the strategy of 
hadronic models, agreed on this number. 

A future experiment, to be conducted at Fermilab, will measure the anomalous magnetic 
moment of the muon $a_\mu$ to a precision of $0.14\,\textrm{ppm}$~\cite{E989exper}. 
It is thus mandatory to work towards getting the LBL contribution under sufficient 
control. For this undertaking we require mature non-perturbative methods that are 
well-rooted in QCD, such as DSE's and lattice QCD. We believe that a promising way 
for the future is to combine these methods in a complementary fashion.

In previous works we have determined important parts of the hadronic LBL 
contributions such as pseudoscalar meson exchange and non-transverse contributions
to the quark-loop part of LBL \cite{Fischer:2010iz,Goecke:2010if}. What has been left 
out so far are the transverse structures of the quark-photon coupling due to numerical 
complexity. From vector-meson dominance models, however, these contributions are 
believed to be sizable and negative and therefore lead to a substantial overall 
reduction of the LBL contribution to the muons anomalous magnetic moment. A complete
approach towards LBL therefore has to include these contributions explicitly. In this
work we provide a further step into this direction. The main focus of this article is,
however, on the comparison between our approach and models like the ENJL or chiral 
quark model. We will argue that such models provide simple tools to give important 
qualitative insights into the significance of different contributions to LBL. However, 
in order to provide more precise quantitative results, more elaborate approaches 
that take into account the full momentum dependence of dressing functions are mandatory.  

This paper is organized as follows. First we briefly introduce the quantity under 
question and give the Dyson-Schwinger 
Equations (DSE) and the particular truncation used in this work, together with some 
notation, in section~\ref{sec:DSEBasics}. 
In section~\ref{sec:ContentofLBL} we 
introduce the hadronic four-point function that lies at the heart of the LBL 
contribution and define it in terms of Green's functions. The main 
body of this work, the quark loop contribution to LBL, is discussed at length in 
section~\ref{sec:QuarkLoopResultAndDiscussion}. Here we make comparison between 
the DSE and the ENJL approach in order to highlight similarities and differences. 
This is followed by a detailed discussion of our numerical results in 
section~\ref{sec:Results}, where the main focus is on the influence of momentum 
dependent dressing functions and the different structures in the self-consistent 
quark photon vertex that are a vital part of our calculation. Details concerning 
the quark-photon vertex and the derivation of the hadronic four-point tensor in 
the present truncation are given in the appendix.
%
%
\section{Basics}
\label{sec:DSEBasics}
To obtain the LBL contribution to the muon anomalous magnetic moment 
$a^\mathrm{LBL}_\mu$ one must consider its contribution to the 
muon-photon vertex shown in Fig.~\ref{fig:LBLContribution}.
On the muon mass-shell this vertex can be decomposed as
\begin{align}
   \parbox{1cm}{\includegraphics[width=0.08\textwidth]{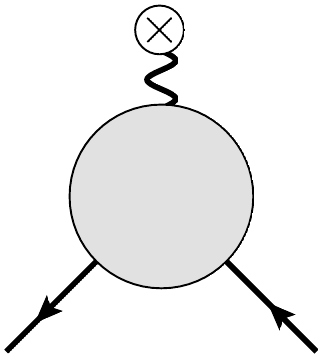}}\quad
   &=\bar{u}(p\prime)
   \left[F_1(q^2)\gamma_\alpha+iF_2(q^2)\frac{\sigma_{\alpha\beta}q^\beta}{2 m_\mu}\right]u(p),
  \label{eqn:MuonPhotonVertexDecomposition}
\end{align}
where $p$ and $p^\prime$ are the muon momenta, $q$ is the photon
momentum and $\sigma_{\alpha\beta}=\frac{i}{2}[\gamma_\alpha,\gamma_\beta]$. The 
anomalous magnetic moment is defined as
\begin{align}
  a_\mu = \frac{g-2}{2}=F_2(0),
  \label{eqn:DefOfAnomaly}
\end{align}
which is obtained from Eq.~(\ref{eqn:MuonPhotonVertexDecomposition})
in the limit of vanishing photon momentum, $q^2$. Here we will use the 
technique advocated in Ref.~\cite{Aldins:1970id} which simplifies the
numerics by ensuring that all integrals are explicitly finite, see
Ref.~\cite{Goecke:2010if} for details.

In the following we introduce the Dyson-Schwinger equations for the quark propagator
and the quark-photon vertex, together with the truncation scheme used in our
calculations.
The dressed quark propagator is given by
\begin{equation}
	S^{-1}(p) = Z_f^{-1}(p^2) \left(  -i\sh{p} + M(p^2) \right)\,\,,
	\label{eqn:inverse_quark_propagator}
\end{equation}
where $Z_f(p^2)$ is the quark wave-function renormalisation and $M(p^2)$
is the quark mass function. These scalar functions are obtained
as solutions to the quark DSE,
\begin{equation}
	S^{-1}(p)=Z_2 S_0^{-1} + g^2 Z_{1f} \frac{4}{3}\!\int\! 
	\overline{dk} \gamma^\mu S(k) \Gamma^\nu(k,p) D_{\mu\nu}(q)\,,
	\label{eqn:quark_DSE}
\end{equation}
where $\overline{dk}=d^4k/(2\pi)^4$ and $q=k-p$ is the gluon's momentum. The bare 
inverse quark propagator is $S_0^{-1}(p) = - i\sh{p} + m_0$. This bare mass is 
related to the renormalized one by $Z_2 m_0 = Z_2 Z_m m_q$, with $Z_2$ and $Z_m$ 
the wave-function and quark-mass renormalisation constants. The renormalisation 
constant for the quark-gluon vertex is $Z_{1f}$. To solve Eq.~(\ref{eqn:quark_DSE}) 
we need the gluon propagator $D_{\mu\nu}(q)$ and the quark-gluon vertex $\Gamma^\nu(k,p)$.

The quark-gluon interaction that appears in the DSE for the quark reads:
\begin{equation}\label{DSEkernel}
  Z_{1f}\,\frac{g^2}{4\pi}\,D_{\mu\nu}(q)\,\Gamma_\nu(k,p) \,.
\end{equation}
In Landau gauge, $D_{\mu\nu} = T_{\mu\nu}(q) Z(q^2)/q^2$ where the transverse 
projector is $T_{\mu\nu}(q) = \delta_{\mu\nu} - q_\mu q_\nu/q^2$. The quark-gluon 
vertex $\Gamma_\nu(k,p)$ can be decomposed into twelve Dirac covariants. However, 
we will employ the rainbow-ladder (RL) truncation, which requires that we replace 
the complicated structure of the quark-gluon vertex with just its $\gamma_\mu$ 
component. Hence Eq.~(\ref{DSEkernel}) becomes
\begin{equation}\label{eqn:trunc2}
Z_{1f}\,\frac{g^2}{4\pi}\,
T_{\mu\nu}(q) \,\frac{Z(q^2)}{q^2}\, \Lambda(q^2)
\gamma_\nu\,\,,
\end{equation}
where $\Lambda(q^2)$ is the non-perturbative dressing of the $\gamma_\nu$ part 
of the quark-gluon vertex, restricted to depend only on the exchanged gluon 
momentum. 
Combining all scalar dressings into one effective running coupling, $\alpha_{\rm eff}(q^2)$ we have
\begin{equation}
Z_{1f}\,\frac{g^2}{4\pi} \,D_{\mu\nu}(q) \,\Gamma_\nu(k,p)
  = Z_2^2 \, T_{\mu\nu}(q) \,\frac{\alpha_{\rm eff}(q^2)}{q^2}\,\gamma_\nu\,,
\end{equation}
where $\alpha_{\rm eff}$ is a renormalization group invariant. 
The factor $Z_2^2$ ensures multiplicative renormalizability.

In contemporary Dyson-Schwinger studies one employs the Bethe-Salpeter equations to
study bound states of two particles, with the interaction described by a two-body kernel.
To provide a realistic description of pseudoscalar mesons one requires that the
dynamical breaking of chiral symmetry is encoded into the truncation.
The symmetry-preserving two-body kernel corresponding to this yields
the `ladder' part of RL. For simplicity we quote this in terms of
$\alpha_{\rm eff}$,
\begin{equation}
	K_{rs,tu}(q)= 4\pi \,Z_2^2 \,\frac{\alpha_{\rm eff}(q^2)}{q^2}\,
	T_{\mu\nu}(q)\,\left[\gamma^\mu\right]_{rt} \left[\gamma^\nu\right]_{us}\,\,.
	\label{eqn:ladder}
\end{equation}
Note that it is possible to employ a beyond-RL truncation here, see 
Refs.~\cite{Fischer:2008wy,Fischer:2009jm,Alkofer:2008et,Chang:2009zb,Bashir:2012fs}.
In practice such an extension complicates the numerics considerably 
and is therefore not yet viable in the context of $g-2$.

We will employ an effective interaction called the Maris-Tandy (MT) 
model which has much phenomenological success for pseudoscalar and 
vector meson masses, decay constants and form factors 
~\cite{Maris:1997tm,Maris:1999nt,Maris:1999bh,Jarecke:2002xd,Maris:2002mz}. 
Success in the meson sector has led to its widespread use in the calculation of 
baryon properties 
\cite{Eichmann:2009qa,Eichmann:2011vu,Eichmann:2011pv,SanchisAlepuz:2011jn}. 
This effective running coupling is given by
\begin{align}
	\alpha_{\rm eff}(q^2) &=
      	 \pi\frac{D}{\omega^2}\,x^2 \, e^{-x}+\frac{2\pi\gamma_m \big(1-e^{-y}\big)}{\log\,[e^2-1+(1+z)^2]}\,, \\
       x &= q^2/\omega^2\,,        \quad
       y = q^2/\Lambda_{t}^2\,,   \quad
       z = q^2/\Lambda_{QCD}^2\,, \nonumber
\end{align}
and features a Gaussian distribution in the infrared that provides dynamical 
chiral symmetry breaking. It is characterized by an energy scale 
$\left(\omega D\right)^{1/3}=0.72$~GeV, fixed to give the pion decay constant, 
and we choose $\omega =0.4$~GeV. The second part reproduces the one-loop 
running coupling at large perturbative momenta. It includes the anomalous 
dimension $\gamma_m=12/(11N_C-2N_f)$ of the quark propagator, and we use 
$\gamma_m=12/25$, $\Lambda_{QCD}=0.234$ GeV and $\Lambda_t=1$ GeV. Note that we 
also employ a Pauli-Villars like regulator with a mass scale of $316$~GeV.
We focus here on the two lightest quarks whose mass at $\mu=19$~GeV is $3.7$~MeV.

The equation for the quark-photon vertex can be written in the form of an inhomogeneous Bethe-Salpeter equation in rainbow-ladder truncation, dependent upon the same two-body ladder kernel Eq.~(\ref{eqn:ladder})
\begin{align}
  [\Gamma_\mu(P,k)]_{rs}=& Z_1 \gamma_\mu \label{eqn:LadderQEDVertexBSE} \\
  - Z_2^2\frac{4}{3}&\int \overline{dq}
  \,[S(q_+)\Gamma_\mu(P,q) S(q_-)]_{ut}K_{tu,rs}(k-q) \,, \nonumber  
\end{align}
where $\overline{dq}=d^4q/(2\pi)^4$, $P$ is the outgoing photon momentum, 
$q_\pm = q \pm P/2$, and $Z_1$ (by the Ward identity $Z_1=Z_2$) is the renormalization constant of the 
quark-photon vertex. The use of rainbow-ladder here ensures that the 
important vector and axial-vector Ward-Takahashi identities (WTIs) 
hold~\cite{Maris:1997hd,Munczek:1994zz}.

For our purposes it is reassuring to have a description that is able to reproduce
a number of observables whilst at the same time being sufficiently simple that we
can define unambiguously the hadronic four-point function.
Furthermore, this approach describes at the same time both the perturbative and 
non-perturbative regime. This unification of IR and UV scales is very important 
for the problem of LBL and is not shared by effective low-energy descriptions 
of the strong interaction. The importance of this feature will be elaborated 
in section \ref{sec:QuarkLoopResultAndDiscussion}.


\section{Content of the hadronic four-point function.}
\label{sec:ContentofLBL}
Now we turn to the structure of the hadronic photon four-point function
$\Pi_{\mu\nu\alpha\beta}$, central to the computation of the hadronic LBL 
contributions. 
Using the rainbow-ladder truncation as introduced above, this object can 
be written exactly in the form (see Appendix~\ref{appB})
\begin{align}
  \begin{array}{c}
  	\includegraphics[width=0.84\columnwidth]{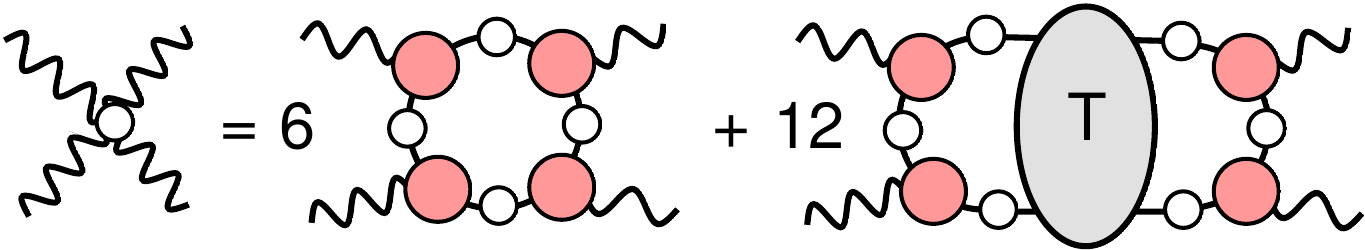}
  \end{array}\,,
  \label{eqn:Hadr4PointFunctionDecomposition}
\end{align}
where the factors $6$ and $12$ indicate the number of permutations of the 
diagrams that one must consider.
We see that there are two classes of diagrams:

\begin{itemize}
\item The second
class of diagram contains the T-matrix that describes all kinds of quark-antiquark
interactions including the dynamical propagation of mesons. In our earlier work,
the T-matrix was approximated by pseudoscalar meson 
exchange~\cite{Fischer:2010iz,Goecke:2010if} in reasonable agreement with 
low energy effective models
\cite{Bijnens:1995xf,Bijnens:2001cq,Hayakawa:1996ki,Hayakawa:1997rq,Hayakawa:2001bb,
Knecht:2001qf,Melnikov:2003xd,Nyffeler:2009uw,Prades:2009tw,Dorokhov:2008pw,Dorokhov:2011zf,Greynat:2012ww}.

\item The first diagram, which we refer to as the quark-loop topology, will constitute
the main focus of this work. This object is composed entirely of fully dressed 
quark propagators and quark-photon vertices. In our previous 
publications~\cite{Fischer:2010iz,Goecke:2010if} we were not in a position to
employ here the full quark-photon vertex as described by
Eq.~(\ref{eqn:LadderQEDVertexBSE}) due to the numerical complexity. Instead, we
were limited to the Ball-Chiu construction~\cite{Ball:1980ay} that fixes the 
first four components of Eq.~(\ref{eqn:QPVMarisTandyBasis}) exactly in terms
of the quark dressing functions, see Eq.~(\ref{eqn:BCVertex}).
\end{itemize}
In this work we will investigate the leading transverse structure
that, amongst other things, dynamically yield the picture of vector meson
dominance (VMD)~\cite{Maris:1999bh}. This is the case since these structures
couple to the vector meson channel and hence one finds time-like poles
corresponding to bound states. The leading component will be extracted and
compared with its ENJL equivalent. Note that we will limit the considerations
to the case of two degenerate flavors for simplicity.
The contributions of strange and charm quarks are only
included for our best estimate of the LBL contribution in section \ref{best}.

Note that there are contributions in the four-point function
that are not accounted for in the representation of \Eq{eqn:Hadr4PointFunctionDecomposition}.
These include unquenching effects due to internal quark lines that can be connected to the 
dynamical back-coupling of hadronic degrees of freedom~\cite{Fischer:2007ze}. In effective 
descriptions such contributions show up as pion-loops that organise themselves 
into a counting scheme within chiral perturbation theory. 
These contributions are typically considered to be sub-leading. A recent investigation,
however, finds that next-to-leading order contributions might be more important than expected
due to the somewhat accidental smallness of the leading terms \cite{Engel:2012xb}.


\section{Quark loop: comparing DSE to ENJL} 
\label{sec:QuarkLoopResultAndDiscussion}

\begin{figure}[b]
  \begin{center}
     \includegraphics[width=0.2\textwidth]{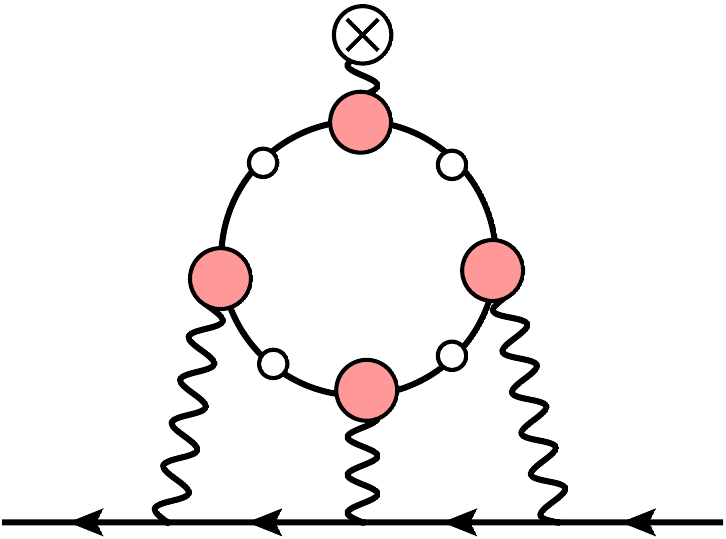} 
  \end{center}
  \caption{The quark loop contribution to the muon $g-2$. The quarks
  and vertices are dressed according to Eqs.~(\ref{eqn:quark_DSE},\ref{eqn:LadderQEDVertexBSE}).}
  \label{fig:LBLQLcontribution}
\end{figure}
In principle we have all of the ingredients at our disposal, see 
Eqs.~(\ref{eqn:MuonPhotonVertexDecomposition},\ref{eqn:DefOfAnomaly})
to calculate the quark-loop contribution to LBL shown in Fig. \ref{fig:LBLQLcontribution}.
We note that this
involves a three-loop integration over fully dressed quark propagators
and quark-photon vertices which poses a considerable numerical
challenge. Thus, to err on the side of caution we present a step-wise 
investigation of going beyond the tree-level approximation reported 
in~\cite{Fischer:2010iz}.

Additionally, it is useful to put our calculation in perspective so that 
a greater understanding of our approach can be conveyed.  To this end, we 
will present a comparison between the DSE approach and the, in many ways 
similar, ENJL model \cite{Bijnens:1995ww,Bijnens:1995ww}. There are of 
course subtle differences between the two, and we will here point out, 
contrast, and discuss the consequences of each. This we further cement 
by testing ENJL-inspired vertices within our approach to explicate their 
kinematical differences.

\subsection{The ENJL perspective}

The inverse quark propagator in the ENJL model is just
\begin{align}
  S^{-1}_{\mathrm{ENJL}}(p) = -i\, \sh{p}+M\,,
  \label{eqn:NJLQuark}
\end{align}
with $p$ a Euclidean momentum and $M$ the constituent quark mass. Note
that the wave function renormalisation is just unity and the mass function
is independent of momentum. 

The quark-photon vertex is given here by a bubble sum
\begin{align}
  \begin{array}{c}
  \includegraphics[width=0.35\textwidth]{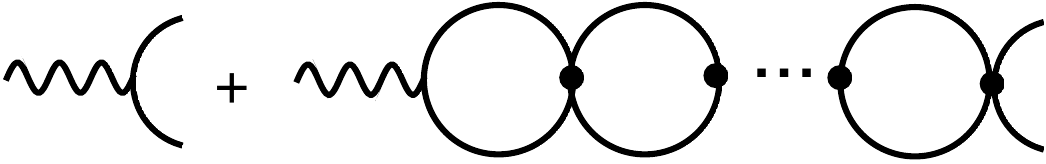}
  \end{array}
  \, , \label{eqn:NJLBubbleSum}
\end{align}
see Refs.~\cite{Bijnens:1995xf,Bijnens:1994ey} for details. 
Owing to the simple structure of the effective four-quark vertices, 
the contributions can be resummed as a simple geometric series of 
one-loop amplitudes $\sum_n \mathrm{Bubble}^n = 1/(1-\mathrm{Bubble})$. 
As a result, the quark-photon vertex depends only on the photon momentum 
$Q^2$ and not on the relative momentum of the quarks\footnote{ 
Note that a similar simplification happens for the four-quark T-matrix, 
which essentially reduces to a bubble sum sandwiched between two pairs 
of quark legs. The resulting mesons are point-like objects.}.

Thus we have
\begin{align}
  \Gamma_\mu^\mathrm{ENJL}(Q) = \gamma_\mu - g_\mathrm{ENJL} \Pi_{\mu\nu}(Q) \gamma_\nu\,,
  \label{eqn:NJLVertex}
\end{align}
where $g_\mathrm{ENJL}$ is derived from standard NJL couplings. 

The bubble sum $\Pi_{\mu\nu}(Q)$ has the transverse structure 
\begin{align}
  \Pi_{\mu\nu}(Q^2) = \left( Q^2\delta_{\mu\nu}-Q_\mu Q_\nu \right) \Pi^\T(Q^2)\,,
  \label{eqn:NJLMesonPropagatorDecomposition}
\end{align}
since a potential longitudinal piece $\Pi^\mathrm{L}(Q^2)$ vanishes for  
identical quark masses~\cite{Bijnens:1994ey}.

According to Ref.~\cite{Bijnens:1994ey} the transverse part in the VMD limit
has the form
\begin{align}
   \Pi^\T(Q^2)= \frac{2 f_V^2 M_V^2}{M_V+Q^2}\,, 
  \label{eqn:TransverseNJLMesonPropagator}
\end{align}
where the momentum dependence of the mass function $M_V$ and decay 
constant $f_V$ are neglected, which is reported to be a rather good
approximation to the momentum dependent case. In the limit we consider
here, $2f_V^2M_V^2= 1/g_{\mathrm{ENJL}}$ and so
\begin{align}
  \Gamma_\mu^\mathrm{ENJL} = \gamma_\mu - \gamma^T_\mu \frac{Q^2}{Q^2+M_V^2}\,,
  \label{eqn:NJLVertexLT}
\end{align}
where $\gamma_\mu^{T}=\gamma_\nu\,T_{\mu\nu}(Q)$ and $M_V$ is identified with
the mass of the $\rho$-meson. Together with the quark propagator \Eq{eqn:NJLQuark}
the vertex satisfies the vector Ward-Takahashi identity (WTI), Eq.~(\ref{eqn:QEDWTI}).
Note that the vertex in \Eq{eqn:NJLVertexLT} can be written as $\sim \gamma_\mu M_V^2/(Q^2+M_V^2)$
if the quark loop is transverse. This explicates why a suppression is found
for VMD inspired transverse dressings compared to a bare vertex $\gamma_\mu$.

\subsection{The DSE perspective}
In the DSE approach, the inverse quark propagator has the following covariant
decomposition
\begin{align}
  S^{-1}_{\mathrm{DSE}}(p) =  Z_f^{-1}(p^2)\,(-i\, \sh{p}+M(p^2))\,,
  \label{eqn:DSEQuarkProp}
\end{align}
where in contrast to Eq.~(\ref{eqn:NJLQuark}) we have a 
momentum dependent wave function $Z_f(p^2)$ and mass function 
$M(p^2)$. 

In the rainbow-ladder truncation employed here, the quark-photon vertex is
a dressed ladder-resummation of effective gluons
\begin{align}
  \begin{array}{c}
  \includegraphics[width=0.82\columnwidth]{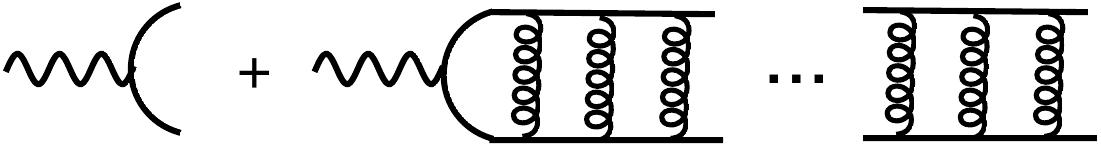}
  \end{array}\,.
  \label{eqn:DSELAdderSum}
\end{align}
If we replace the gluon exchange by a momentum independent contact interaction 
we arrive at a picture similar to the ENJL model above. Keeping the
exchange as is, the calculation is somewhat more involved as we can no longer
have a simple geometric series.

The full quark-photon vertex is given by 
\begin{align}
  \Gamma_\mu(Q,k) = \sum_{i=1}^{4} \lambda^{(i)}L_\mu^{(i)}(Q,k) + \sum_{i=1}^{8} \tau^{(i)}T_\mu^{(i)} (Q,k)\,,
  \label{eqn:VertexDecomposition}
\end{align}
with $\lambda^{(i)}=\lambda^{(i)}(Q^2,k^2,Q\cdot k)$ and
$\tau^{(i)}=\tau^{(i)}(Q^2,k^2,Q\cdot k)$ the longitudinal and transverse 
scalar coefficients respectively that correspond to the basis elements given 
in Eq.~(\ref{eqn:QPVMarisTandyBasis}). Additionally $\Gamma^{\mathrm{BC}}_\mu =\sum_{i=1}^{4} \lambda^{(i)}L_\mu^{(i)}(Q,k)$ defines the Ball-Chiu vertex with the coefficients fixed
by the vector WTI \cite{Ball:1980ay}.  
Our main interest here lies in the eight transverse components $T_\mu^{(i)} (Q,k)$
which couple to vector bound-states. For the sake of comparison with the ENJL
model, we will take only the leading transverse component $T_\mu^{(1)}(Q,k)=\gamma_\mu^T$ under
consideration in this work.

While we are mainly working with the full numerical result for the quark-photon vertex
it is sometimes useful to also have a simple analytical form at hand, which captures
the main features of the numerical solution.
An approximate form for the leading component of the transverse part has been given in
Ref.~\cite{Maris:1999bh} which depends on both the relative and total momenta
of the vertex
\begin{align}
  \Gamma_\mu(Q,k) \simeq\Gamma_\mu^\mathrm{BC}\!-\! 
          \gamma_\mu^\T \frac{\omega^4N_V}{\omega^4+k^4}
	  \frac{f_V}{M_V}\frac{Q^2}{Q^2+M_V^2}\, e^{-\alpha(Q^2+M_V^2)}.
  \label{eqn:QEDVertexFitToLeadingTransverse}
\end{align}
This form has been fitted to the full numerical solution for the quark-photon 
vertex obtained from its BSE. Here, $\omega$ and $\alpha$ describe the suppression 
of the amplitude for large relative and total momentum respectively. We find 
reasonable agreement with the numerical solution with $\omega = 0.66\,\mbox{GeV}$, 
$\alpha=0.15$ and $N_V f_V/M_V = 0.152$, see Fig. \ref{fig:TransverseVertexDSENJL}.
Additionally, $N_V$ is a normalization factor. We consider this formula to be the
VMD limit of the DSE quark-photon vertex. In addition to this BSE inspired fit, 
we will also employ the $T^{(1)}_\mu$ component of the vertex as extracted from 
the full calculation of the quark-photon vertex BSE.


\subsection{Differences between DSE and ENJL}
One of the key differences between the ENJL and DSE approaches is that in the former we
have a contact interaction, whilst the latter features an interaction that 
features momentum exchange. This has far-reaching consequences as we discuss below. 

\begin{figure}[b]
  \begin{center}
    \includegraphics[width=0.33\textwidth,angle=-90]{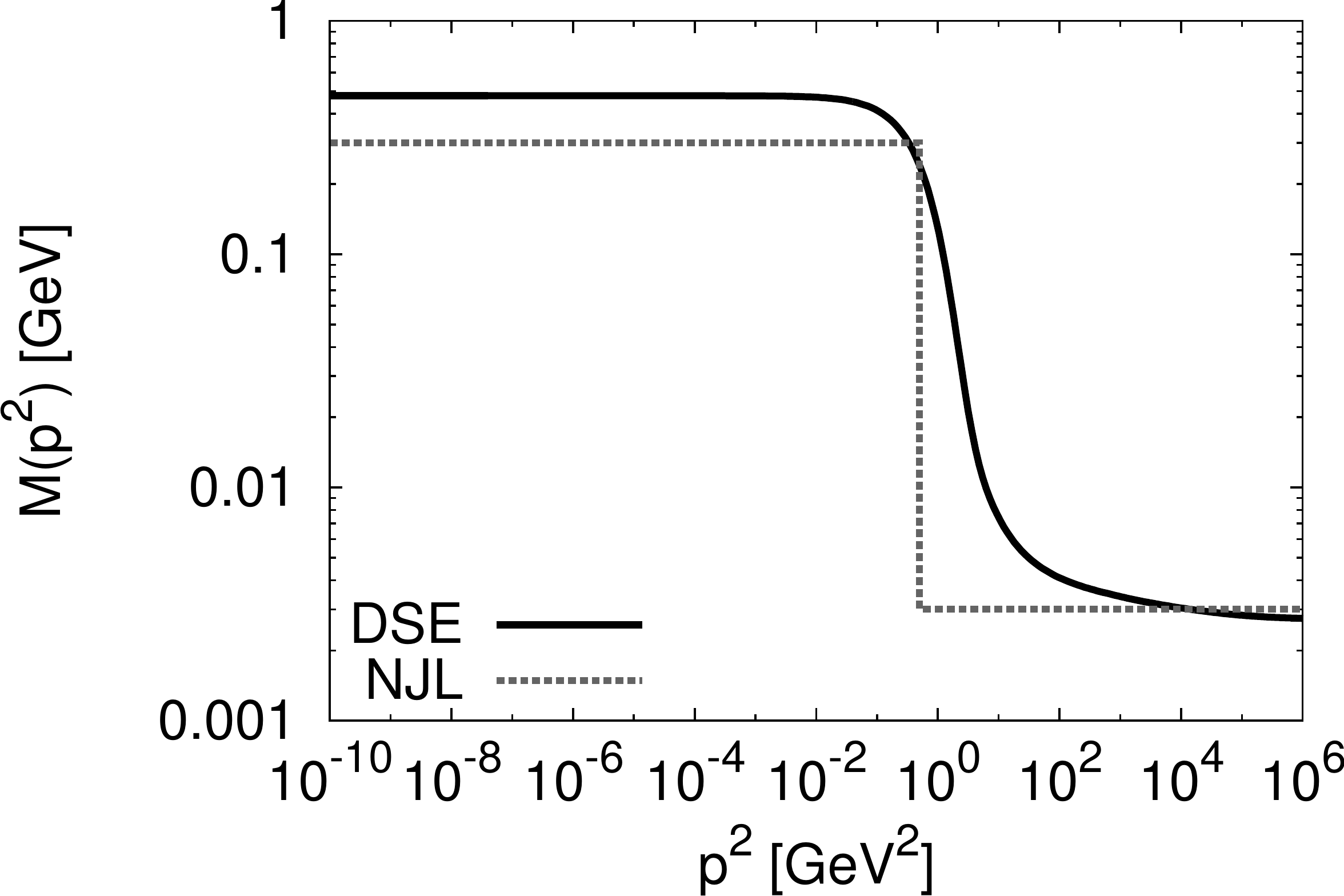}\vspace*{6mm}
    \includegraphics[width=0.33\textwidth,angle=-90]{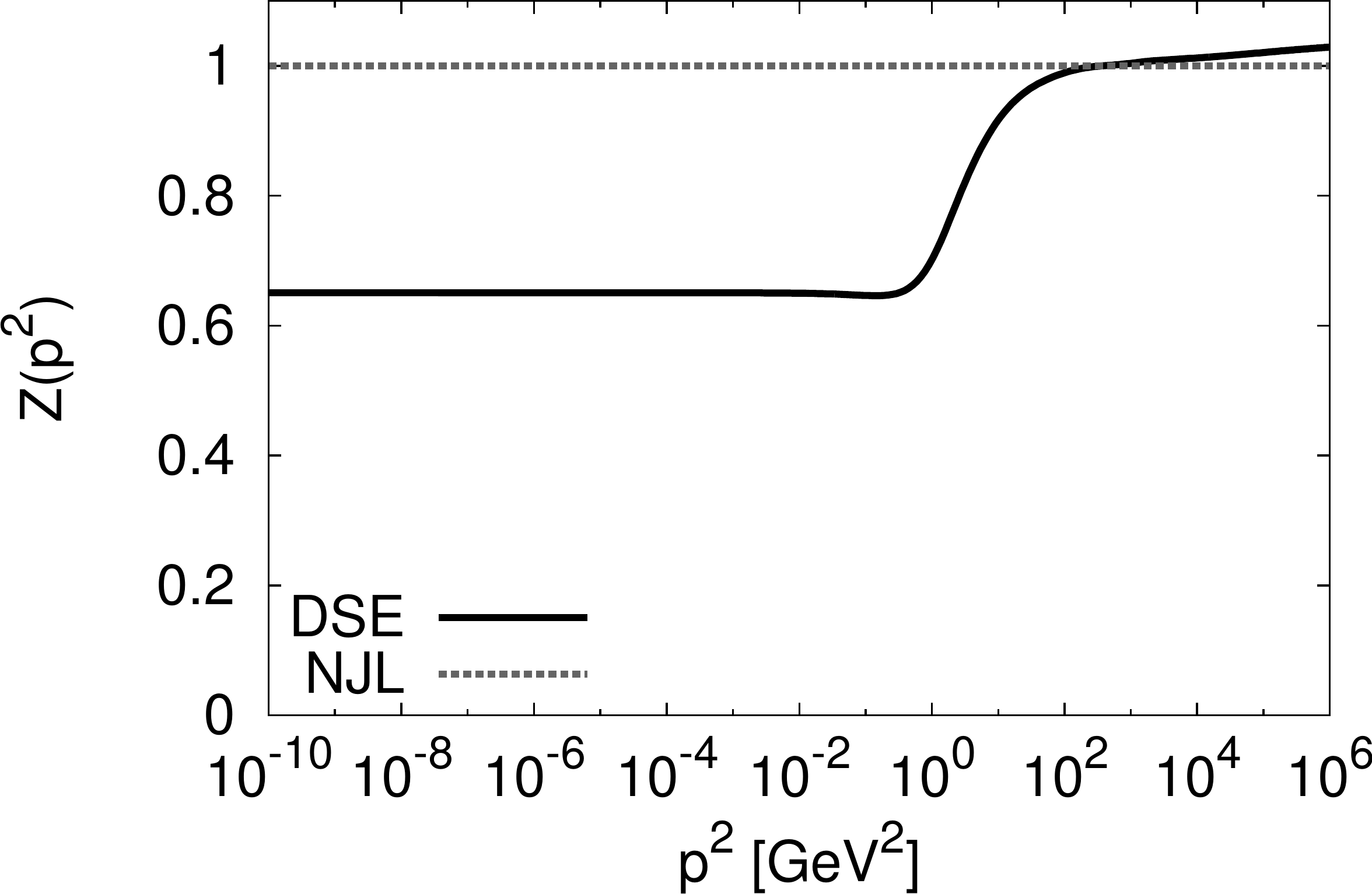}
  \end{center}
  \caption{Comparison of the mass function $M$ and the wave function
  $Z_f$ for DSE and ENJL quarks.}
  \label{fig:QuarkMandZdseANDnjl}
\end{figure}

Firstly we look at the differences in the quark propagator, see Fig.~\ref{fig:QuarkMandZdseANDnjl}.
In the ENJL model we have $Z_f(p^2)=1$ and $M(p^2)=M_{\textrm{const}}$ 
as opposed to the fully momentum dependent functions from the DSE. 
For the quark mass function, we see that in the DSE it saturates
in the IR at about $M(0)\approx 450\,$MeV and continuously connects to its 
perturbative running at large momenta. The ENJL model, in contrast features a
constituent-like quark mass of $\sim 300\,$MeV at all scales up until the model
cut-off $\sim 1$~GeV.
For the quark wave function renormalisation, we see that for a large momentum range
$Z_f(p^2)<1$ which constitutes a suppression of the quark propagator with respect 
to the constant $Z$ of the ENJL model.
This can have several
consequences for the quark-loop contribution to $g-2$, Fig.~\ref{fig:LBLQLcontribution}, 
where there are four quark propagators. With $Z_f(p^2)<1$ in the DSE approach, we may expect 
a suppression of the contribution by a factor $Z_f(s)^4$ for some representative, `average' 
momentum scale $s$. On the other hand, the momentum dependent quark-mass function allows 
for quark masses smaller than $M(p^2=0)$ to be probed. Naively, this leads to an enhancement 
of the contribution. This may be equivalent to using a momentum-independent quark mass $M(s)$ 
where $s$ is some representative momentum scale. This `effective' mass may be surprisingly 
small and, indeed, such small quark masses have been observed to be necessary in several 
models \cite{Dorokhov:2008pw,Greynat:2012ww,Boughezal:2011vw}. We come back to this point below.

If we now compare the ENJL vertex, Eq.~(\ref{eqn:NJLVertexLT}) and its DSE equivalent,
Eq.~(\ref{eqn:QEDVertexFitToLeadingTransverse}) we see that there is a similar structure. That
is, there is a part dictated by the WTI (gauge part) and a part that represents the VMD physics of the
transverse vertex. Whilst in the case of the DSE we have functions that depend on both the
photon and relative quark momenta, in the case of the ENJL model we have trivial momentum
dependence for the gauge-part, and reduced momentum dependence for the transverse part.

\begin{figure}[b]
  \begin{center}
    \includegraphics[width=0.65\columnwidth,angle=-90]{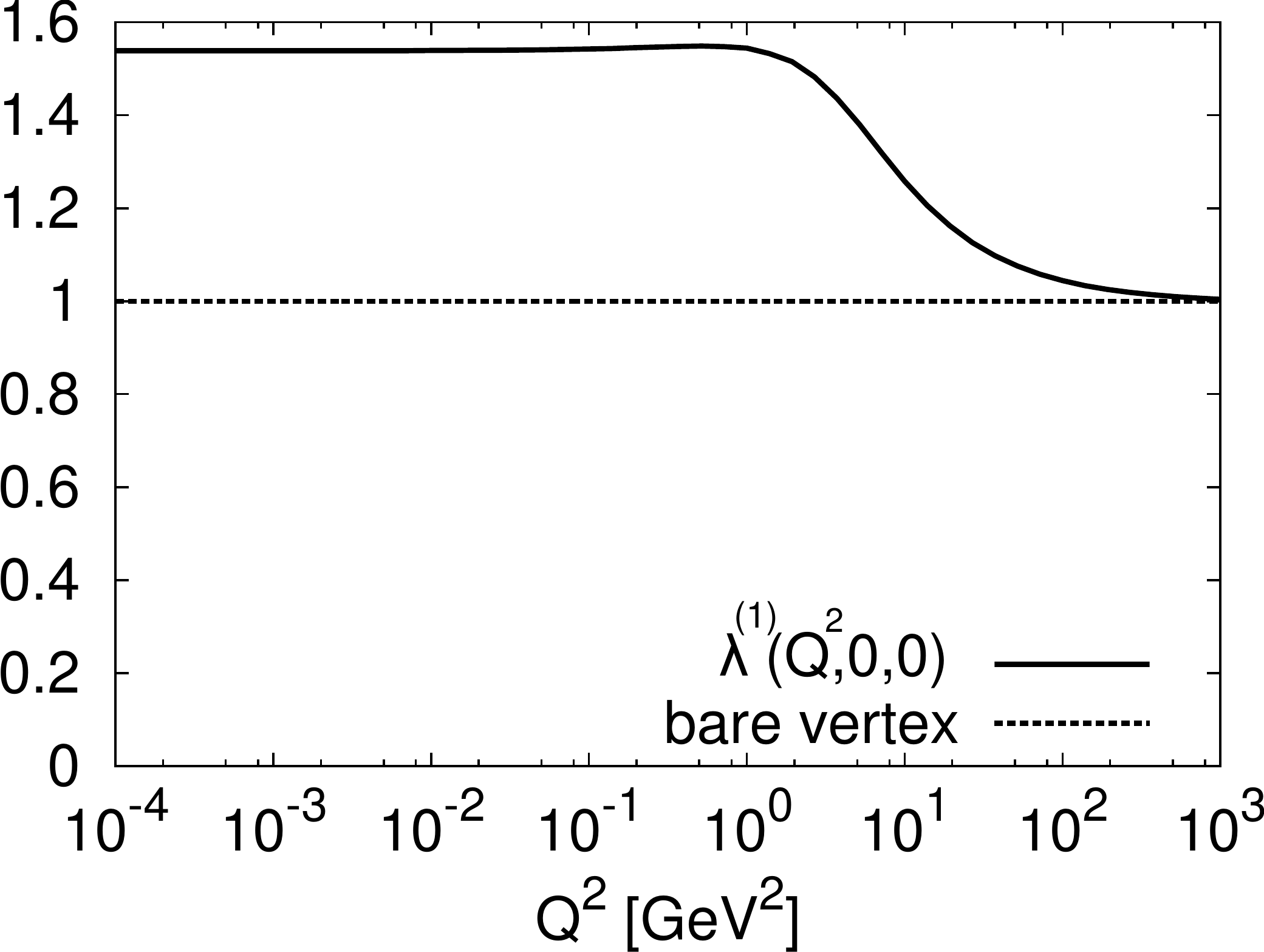}
    \includegraphics[width=0.65\columnwidth,angle=-90]{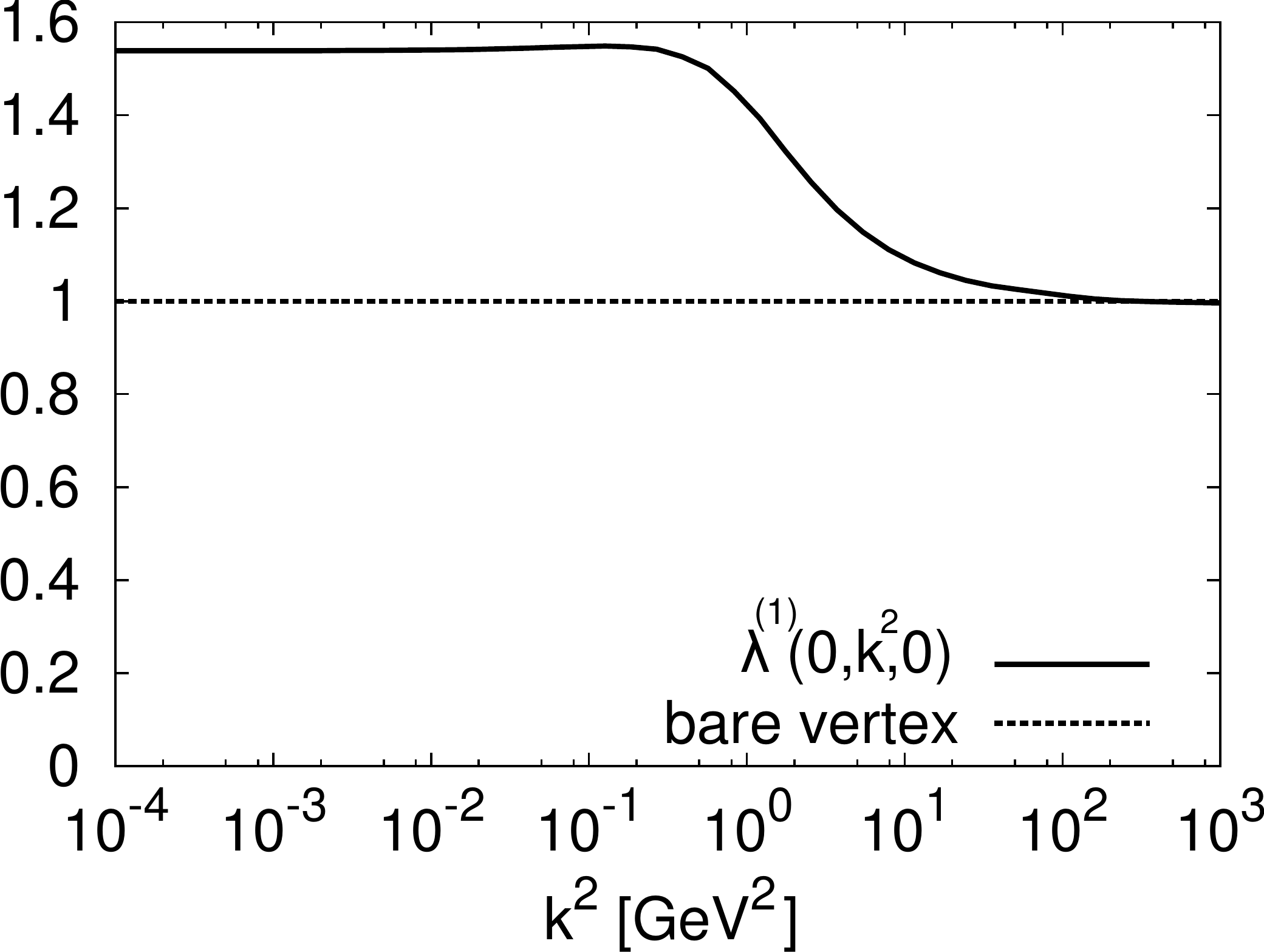}
  \end{center}
  \caption{The leading $\lambda^{(1)}$ component of the quark-photon vertex constrained by the Ward-Takahashi identity of  Eq.~(\ref{eqn:QEDWTI}) and given in Eq~(\ref{eqn:1BCVertex}). We show two slices as
  a function of the total and relative momenta $Q^2$ (with $k=0$) and $k^2$ 
  (with $Q=0$), respectively. Note that the
  constant dressing corresponds to the ENJL model.}
  \label{fig:1BCVertexDSENJL}
\end{figure}

First, let us discuss the gauge part of the quark-photon vertices. The
leading coefficient of $\gamma_\mu$ has the form
\begin{align}
  \lambda^{(1)}(Q^2,k^2,k\cdot Q) = \frac{1}{2}\left( \frac{1}{Z_f(k_+^2)}+\frac{1}{Z_f(k_-^2)}\right)\,,
  \label{eqn:1BCVertex}
\end{align}
where $k_\pm = k \pm Q/2$. Based on the difference in behavior of the 
quark propagators, we see for the ENJL model that $\lambda_1=1$, 
whilst for the DSE $\lambda_1>1$. Thus, for the DSE we expect the gauge part 
of the vertex to yield an enhancement of the quark-loop $\sim 1/Z_f(s)^4$ as 
to the bare vertex approximation. A comparison of
these components of the vertex is shown in 
Fig.~\ref{fig:1BCVertexDSENJL}.

\begin{figure}[t]
  \begin{center}
    \includegraphics[width=0.35\textwidth,angle=-90]{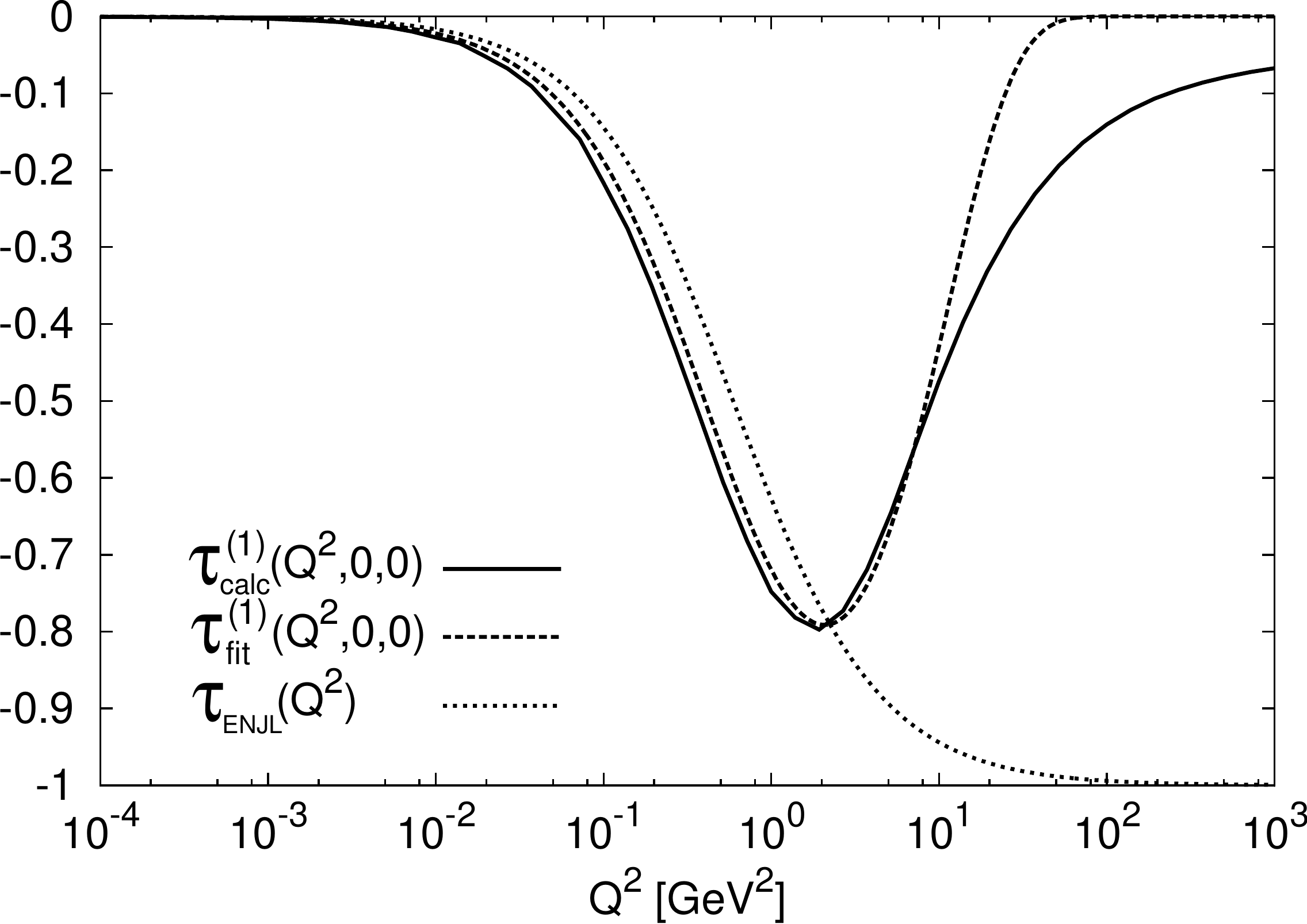}
  \end{center}
  \caption{The dependence of the dominant transverse dressing $\tau^{(1)}$ on 
  the photon momentum $Q^2$ is shown for the explicit solution of the quark-photon
  vertex BSE, Eq.~(\ref{eqn:LadderQEDVertexBSE}), the fit to this given in
  Eq.~(\ref{eqn:QEDVertexFitToLeadingTransverse}), and the transverse part of the ENJL vertex,  
  Eq.~(\ref{eqn:NJLVertexLT}).}
  \label{fig:TransverseVertexDSENJL}
\end{figure}

Now we take a look at the dominant transverse component of the quark-photon
vertex. In Fig.~\ref{fig:TransverseVertexDSENJL} we show the dressing function
as calculated self-consistently from the DSE for the quark-photon vertex, together
with the fit function of Eq.~(\ref{eqn:QEDVertexFitToLeadingTransverse}) and
the equivalent part of the ENJL vertex, Eq.~(\ref{eqn:NJLVertexLT}). We see that
the fit to the DSE, as a function of the total momentem $Q$, behaves very similarly
as the full numerical solution in the dominant region around the scale $M_V=0.77\,$GeV.
Small deviations occur at large momenta due to the exponential fall-off of the fit function. The dressing 
function of the ENJL model has a similar behavior except at large momenta where no 
fall-off is seen and instead it tends to a constant. However, due to the weighting 
of the integrand in the calculation of $a_\mu$ it transpires that such deviations 
at large $Q^2$ are not relevant.

However, differences between the DSE and ENJL approach become readily apparent when one
considers the impact of the relative momentum on the kinematics. In the DSE approach the 
transverse VMD-piece is suppressed for momenta $k^2\gtrsim \omega^2 \approx \Lambda^2_\QCD$, 
an effect which is not present in the ENJL approach due to its contact interaction. 
As a consequence, we expect smaller negative contributions to LBL in the DSE approach
as compared to ENJL. The degree of overestimation of negative contributions in ENJL 
depends upon the kinematical weighting of the integrand, and can be
estimated in our approach by systematically removing the dependence on the relative
momentum in our quark-vertex dressing function. This will be detailed in the next section.


\section{Results}\label{sec:Results}
Here we present a more quantitative analysis of the difference between the ENJL and DSE
approaches, with respect to various approximations that can be made in the calculation
of the quark loop. In order to compare the two approaches we need a means to find the
most important and therefore representative momentum scales. This we achieve by averaging 
over the dressing functions, as weighted by the importance sampling of the VEGAS Monte-Carlo 
\cite{Hahn:2004fe} we use to evaluate the quark-loop contribution to LBL.

For the calculation in the DSE-framework we use an IR-cutoff of $10^{-3}$ GeV and an UV-cutoff
of $10^{3}$ GeV for all three loop variables. No additional splitting into a perturbative
and a nonperturbative region is necessary, since these regions are continuously connected,
see the discussion around Fig.~\ref{fig:QuarkMandZdseANDnjl}. In contrast, this is different
for the ENJL model where the calculation is split into a low and a high-energy part 
\cite{Bijnens:1995xf}. In the former, constituent quarks
Eq.~(\ref{eqn:NJLQuark}) with $M=0.3$ GeV are used together with the vertex construction
Eq.~(\ref{eqn:NJLVertexLT}). This low-energy part is cut off by requiring the photon
momenta to be $q\leq \mu$, where $\mu$ is varied in the range between $0.7$ GeV
and $4$ GeV. The high-energy contribution is approximated by a 
bare quark-loop (bare propagators and bare vertices) where the mass of the quarks
is given by the same scale $\mu$ which is supposed to act as an effective IR-cutoff.
It was found in \cite{Bijnens:1995xf} that the result is rather stable against a variation of $\mu$
in the considered range. The final result is quoted as
\begin{align}
  a_\mu^{\mathrm{LBL,ql, ENJL}}=21(3)\times 10^{-11}
  \label{eqn:ENJLQuarkLoopResult}.
\end{align}
Note that if $\mu=4$ GeV is taken, almost the complete contribution comes from
the low-energy part. While this value is usually considered to be inappropriately 
high for an NJL type model, we find this viewpoint useful since it facilitates 
the comparison to the DSE case, where there is no separation in high and low-energy
parts. We will later see that the contribution Eq.~(\ref{eqn:ENJLQuarkLoopResult})
can be roughly reproduced even with the much higher cutoff we use in our 
calculations\footnote{
Note furthermore that  Eq.~(\ref{eqn:ENJLQuarkLoopResult}) includes strange quarks
which we neglect in this work. These give, however, a few percent correction \cite{Bijnens:1995xf}
such that this detail is not important for the more general analysis we present here.}.
The results presented in this work are obtained within Landau gauge for QED 
($\xi=0$ in the photon propagator apparent in fig. \ref{fig:LBLQLcontribution}), in contrast
to our earlier work \cite{Fischer:2010iz,Goecke:2010if} where we used Feynman gauge for QED ($\xi=1$).
It turns out that all the results presented here are affected only mildly by this choice, as 
can be seen in Table \ref{tab:quarkloopMTAWWcomparison}.

\subsection{Impact of a dynamical $Z_f$ and $M$}
To gauge the impact of momentum dependent quark dressing functions, we compare
several calculations in the bare-vertex approximation,
see Table \ref{tab:LBLResultsBareVertex}. Note that we attach a renormalisation constant
to the bare vertex e.g. $Z_2\gamma_\mu$ to ensure multiplicative renormalisability. 
For a momentum independent quark wave-function we also supplement the quark by a factor
$Z_2^{-1}$ for the same reason. These extra factors cancel in all but the first line
of Table \ref{tab:LBLResultsBareVertex}. The corresponding case of bare vertex with
fully dressed quark thus differs from our earlier results \cite{Fischer:2010iz,Goecke:2010if} by a factor of $Z_2^4\approx0.89$
in addition to the different QED gauge and the smaller number of quark flavours.

\begin{table}[t]
\begin{center}
\begin{tabular}{l|c|c|c}
  Quark dressing 
& $\langle Z_f\rangle$	
& $\sqrt{\langle M^2\rangle} $~[GeV]
& $a_\mu [10^{-11}]$ \\
     \hline
     \hline
   $Z_f$ and $M$ dynamical	& $0.83$	&	$0.14$	&	$47$	\\
   $Z_f=Z_2^{-1}$, $M$ dynamical  	&--		&	$0.18$	&	$102$\\
   $Z_f=Z_2^{-1}$, $M=0.477$  	&--		&	-- &	$22$	\\   
   $Z_f=Z_2^{-1}$, $M=0.300$  	   	&--		&	--   & 	$51$	\\
   $Z_f=Z_2^{-1}$, $M=0.200$    	&--		&   -- 	&	$104$	\\

  \end{tabular}
 \end{center}
    \caption{The quark-loop contribution to hadronic LBL together with effective average values for 
    the $Z_f$ and $M$ functions determined with a bare quark-photon vertex $Z_2\gamma_\mu$. We compare results for 
    dynamical quark dressing functions to those with static ENJL-like equivalents. The quark wave-function
    renormalisation constant $Z_2\approx 0.97$ for the cutoff and renormalisation point used.}
  \label{tab:LBLResultsBareVertex}
\end{table}

Let us first compare the first two lines of Table \ref{tab:LBLResultsBareVertex}. Replacing the dynamical quark wave function
by $Z_f=Z_2^{-1}$ we note an enhancement of the contribution to $a_\mu$ roughly by a factor of 2.
This is in good agreement with the expectation discussed above: since there are four 
quark propagators in the quark loop, the full calculation contains an extra factor of the order 
$\langle Z_f\rangle^4$ with $\langle Z_f \rangle$ the average value of the wave-function 
that is probed. Indeed, one finds $\left< Z_f \right>^4 \simeq (0.83)^4 \sim 0.5$.

To explore the impact of the momentum dependent quark mass function, we keep $Z_f=Z_2^{-1}$ 
fixed and compare several constant values of $M$ to the dynamical case.
For $M=0.477$ GeV, the infrared plateau of the dynamical mass function, we 
obtain less than $1/5$th of the dynamical result (with $Z_f=Z_2^{-1}$). A quark mass of 
$M=0.3$ GeV gives a result commensurate with the calculation of Ref. \cite{Kinoshita:1984it}
which yields $a_\mu^{LBL,ql,N_f=2}=49.1(3.4)\times 10^{-11}$
for the contribution
of bare $u$ and $d$ quarks of the same mass $M=0.3\,$GeV \footnote{This number 
is extracted from Table I of Ref. \cite{Kinoshita:1984it}}.
Only a significantly smaller constant mass $\sim0.2$ GeV close to
the average one probed under the integrand leads to a result that compares with 
the dynamical one. Note, however, that this does not work in general, i.e. the 
fully dynamical result of $47 \times 10^{-11}$ is not reproduced by a static approach
with $Z=0.83$ and $M=0.14$ GeV.

\subsection{Impact of dressed vertices: Gauge Part}
Here we focus upon the gauge part of the vertex. In the case of the ENJL
model this part is just bare since $Z_f=1$. In the DSE approach for the leading 
part, Eq.~(\ref{eqn:1BCVertex}), we give the fully dynamical result in 
Table~\ref{tab:LBLResults1BCVertex}. By comparison with the first value of 
Table \ref{tab:LBLResultsBareVertex} we find an increase by more than a factor of 2
when the vertex dressing is included. Again, this is roughly what one expects, 
since the contribution from the vertices gives a factor $\sim \langle Z_f \rangle^{-4}$. 
Note that the enhancement from the gauge-part of the vertex is comparable to the
suppression due to a non-trivial $Z_f$ in the quark propagator. 

In principle, we should include not only the leading term of the gauge part of 
the vertex but the full Ball-Chiu vertex as fixed by the Ward-Takahashi identity.
However, this is currently not possible due to numerical instabilities which
are currently not under control\footnote{The corresponding results given in 
Ref.~\cite{Goecke:2010if} are presumably not correct as will be detailed in an erratum.}.

\begin{table}[t]

  \centering
  \begin{tabular}{l|c|c}
    Quark Dressing			&	$\langle Z_f\rangle$	 & $a_\mu [10^{-11}]$	\\
    \hline
   $Z_f$ and $M$ dynamical		
   &$0.76$	&	$100$	\\
  \end{tabular}
    \caption{The quark-loop contribution to hadronic LBL with the 1BC dressing and the average of the $Z_f$ function. These are to be compared with the bare-vertex results in the previous table. }
  \label{tab:LBLResults1BCVertex}
\end{table}

\subsection{Impact of dressed vertices: VMD Part}

To better compare with the ENJL model we consider first the case where 
the quark-photon vertex is taken to be of the form 
\begin{align}
   \Gamma_\mu(Q,k) = Z_2\gamma_{\mu} + \tau^{(1)}(Q,k)\; T_\mu^{(1)}(Q,k)\,,
\end{align}
with the leading transverse component $T_\mu^{(1)}(Q,k) = \gamma_\mu^T$, and $\tau^{(1)}(Q,k)$ 
its dressing function. The quark wave function renormalisation constant $Z_2$ is necessary
to maintain multiplicative renormalizability. 
For the transverse dressing function $\tau^{(1)}(Q,k)$ we study three choices: (i)
the ENJL model, Eq.~(\ref{eqn:NJLVertexLT}), which we also supplement by an additional
factor 
\begin{equation}
f(k^2)=1/(1+k^4/0.66^4)
\end{equation}
to simulate a relative momentum dependence, (ii)
the DSE approach with the numerical solution to Eq.~(\ref{eqn:LadderQEDVertexBSE}) and
(iii) the DSE approach with the analytic fit function Eq.~(\ref{eqn:QEDVertexFitToLeadingTransverse})
to the numerical result. Both, (ii) and (iii) already include a relative momentum dependence 
in contrast to the original ENJL approach. Our results are presented in 
Table~\ref{tab:bareANDtransverseVertexResults}.

First of all note that the result where a bare vertex $Z_2\gamma_\mu$ is used
together with the ENJL type transverse part (first line) is quite close to the 
ENJL result shown in Eq. (\ref{eqn:ENJLQuarkLoopResult}). We are thus able to 
reproduce the ENJL result numerically, despite the very different UV cutoff, that 
obviously does not matter much. 
Taking a closer look at Table \ref{tab:bareANDtransverseVertexResults} we see a common pattern.
When the vertices do not depend on the relative
momentum between the two quarks, the contribution is around $a_\mu\sim (14-16)\times 10^{-11}$.
Compared to the corresponding result with bare-vertex of $47\times 10^{-11}$ we therefore 
find a reduction of similar size as the one from $\sim 60\times 10^{-11}$ to $\sim 20\times 10^{-11}$ reported by \cite{Bijnens:1995xf}. We obtain this large suppression in our ENJL type 
calculation as well as in the DSE approach with momentum dependent quark propagators.
However, when we take into account the relative momentum dependence in the full DSE 
calculation and including the additional factor $f(k)$ in the ENJL model we find that 
the suppression due to transverse parts is much reduced. We find results in the range
of $a_\mu\sim (41-46) \times 10^{-11}$. This is at most a reduction of just $\sim 15\%$
and constitutes one of the main results of this work.

\begin{table}[t]

  \centering
  \begin{tabular}{l|c}
	Vertex Dressing	&	
	$a_\mu [10^{-11}]$	\\
	\hline\hline
	$Z_2\gamma_\mu+\gamma^\T_\mu \tau_{\mathrm{ENJL}}$	&
	$14$	\\
	\hline
	$Z_2\gamma_\mu+\gamma^\T_\mu \tau_{\mathrm{ENJL}}f(k^2)$ &
	$45$ 	\\
	\hline\hline
	$Z_2\gamma_\mu+\gamma^\T_\mu \tau^{(1)}_{\textrm{fit}}(k=0)$		&
	$16$	\\
	\hline
	$Z_2\gamma_\mu+\gamma_\mu^\T \tau^{(1)}_{\textrm{fit}}$	&
	$46$	\\
	\hline\hline
	$Z_2\gamma_\mu+\gamma^\T_\mu \tau^{(1)}_{\textrm{calc}}(k=0)$		&
	$14$	\\
	\hline
	$Z_2\gamma_\mu+\gamma_\mu^\T \tau^{(1)}_{\textrm{calc}}$	&
	$41$	
  \end{tabular} 
    \caption{Bare and leading transverse vertex component, with dressing functions 
     from the ENJL model, VMD like fit from DSE/BSE ($\tau^{(1)}_{\textrm{fit}}$), and
     from an explicit calculation of the quark-photon DSE ($\tau^{(1)}_{\textrm{calc}}$).
     Results are shown with, and without the inclusion of a 
     dependence on the relative momentum.}
\label{tab:bareANDtransverseVertexResults}
\vspace*{4mm}
  \centering
  \begin{tabular}{l|c}
	Vertex Dressing	&	
	$a_\mu [10^{-11}]$	\\
	\hline\hline
	$\gamma_\mu \lambda^{(1)}+\gamma^\T_\mu \tau_{\mathrm{ENJL}}$	&	
	$43$	\\
	\hline
	$\gamma_\mu \lambda^{(1)}+\gamma^\T_\mu \tau_{\mathrm{ENJL}} f(k^2)$ &	
	$103$	\\
	\hline\hline
	$\gamma_\mu \lambda^{(1)}+\gamma^\T_\mu \tau_\mathrm{fit}^{(1)}(k=0)$		&
	$43$	\\
	\hline
	$\gamma_\mu \lambda^{(1)}+\gamma_\mu^\T \tau_\mathrm{fit}^{(1)} $	&
	$105$	\\
	\hline\hline
	$\gamma_\mu \lambda^{(1)}+\gamma^\T_\mu \tau_\mathrm{calc}^{(1)} (k=0)$		&
	$41$	\\
	\hline
	$\gamma_\mu \lambda^{(1)}+\gamma^\T_\mu \tau_\mathrm{calc}^{(1)} $	&
	$96$	
  \end{tabular}
    \caption{1BC and leading transverse vertex component, with dressing functions 
     from the ENJL model, VMD like fit from DSE/BSE ($\tau^{(1)}_{\textrm{fit}}$), and
     from an explicit calculation of the quark-photon DSE ($\tau^{(1)}_{\textrm{calc}}$).
     Results are shown with, and without the inclusion of a 
     dependence on the relative momentum.}
  \label{tab:1BCandTransverseVertexResults}
\end{table}

Finally, we give the same comparison with the inclusion of the 
leading $L_1$ dressing $\lambda^{(1)}$ from \Eq{eqn:1BCVertex} given by the WTI, see Table~\ref{tab:1BCandTransverseVertexResults}.
We see that the trend here is very similar as for the bare vertex, except now with 
an enhancement due to the non-trivial dressing function of the gauge part.
This enhancement is of the order of $2-3$. It differs slightly for the cases with and
without relative momentum dependence, which shows that there is also interference
between the different vertex components. Note that the $\lambda^{(1)}$ dressing
is always used with its full kinematics.

\subsection{Best result and electromagnetic gauge invariance}\label{best}

Our most reliable estimate for the contribution from the quark-loop to
the anomalous magnetic moment of the muon is the one obtained with
full dynamics in the quark propagator, the leading gauge-part $\lambda^{(1)}$ 
and the leading transverse part $\tau^{(1)}_{calc}$ of the quark-photon
vertex. For two light quark flavors we obtained 
\begin{align}
  a_\mu^{\mathrm{LBL,L1+T1},N_f=2} = (96 \pm 2) \times 10^{-11},
  \label{eqn:1BCplusLeadingTransverseLBLResult}
\end{align}
where the error is purely numerical. Compared to the corresponding value 
$a_\mu^{\mathrm{LBL,L1},N_f=2} = (100 \pm 2) \times 10^{-11}$
for the
1BC vertex without transverse parts we thus find a suppression of the order
of five percent due to the VMD physics. This is much less than in simple
models. If we additionally include the strange and charm quark contributions 
we arrive at
\begin{align}
  a_\mu^{\mathrm{LBL,L1+T1},N_f=4} = (107 \pm 2) \times 10^{-11},
  \label{eqn:1BCplusLeadingTransverseLBLResultudsc}
\end{align}
which compares to the 1BC $N_f=4$ case (in Landau gauge)
$a_\mu^{\mathrm{LBL,L1},N_f=4} = (111 \pm 2) \times 10^{-11}$.
Note once more that the corresponding result in \cite{Goecke:2010if}
differs slightly due to QED Feynman-gauge, see discussion below.

We are currently working on the further inclusion of the other transverse 
terms $\tau^{(2..8)}$, corresponding results will be presented elsewhere.
The potential impact of these terms is hard to gauge without an explicit 
calculation. Nevertheless, from a systematic point of view, the omission
of these terms is unproblematic. This is different for the non-transverse
part of the vertex. Strictly speaking, the presence of all three Ball-Chiu 
components $\lambda^{(1..3)}$ are necessary to maintain electromagnetic gauge
invariance. As mentioned above, this is currently not possible due to 
severe numerical problems with the terms $\lambda^{(2)}$ and $\lambda^{(3)}$.
We therefore have to gauge the error in the present calculation due to
violations of gauge invariance. This is conveniently done by varying the
QED gauge parameter $\xi$. Results for Feynman and Landau gauge are
shown in table \ref{tab:quarkloopMTAWWcomparison}. The variations with 
$\xi$ are on the two percent level and therefore reassuringly small. 
The insensitivity with respect to the gauge parameter is an indicator 
for the (almost) transversality of the resulting quark-loop part of the
photon four-point function $\Pi_{\mu\nu\alpha\beta}$. 
\begin{table}[t]

  \centering
  \begin{tabular}{c|c|c}
    Interaction & $\xi$  &   $a_\mu^{\mathrm{LBL,QL}}\times 10^{11}$		\\
    \hline\hline
    MT		& $0$ & $96$  \\
    MT		& $1$ & $94$	\\
  \end{tabular}
   \caption{Results for the 1BC+transverse vertex dressings. We compare different
 photon gauge parameters $\xi=\{0,1\}$, i.e. Landau and Feynman gauge.}
   \label{tab:quarkloopMTAWWcomparison}
\end{table}

We emphasize, however, that the smallness of the gauge violations
due to the omission of $\lambda^{(2)}$ and $\lambda^{(3)}$ cannot be taken
as an indication that these terms will not contribute much to the 
physical, transverse part of the photon four-point functions. As already
mentioned above, our previous calculation of these contributions in 
Ref.~\cite{Goecke:2010if} is presumably not correct and needs to be
thoroughly reinvestigated. This will be done in future work.


\section{Summary and conclusion}
In this work we investigated in detail key similarities and differences between 
the ENJL and DSE approaches. Whereas the ENJL model features a contact interaction 
giving rise to a trivial quark wave-function renormalisation and a constant quark 
mass function, in the DSE approach these are both momentum dependent. The same is
true for the gauge part of the vertex, that is determined by a Ward-Takahashi Identity.
Whereas in the ENJL model this part is
trivial, the corresponding Ball-Chiu terms in the DSE approach are nontrivial
and momentum dependent. In both approaches there are transverse parts in the vertex
which are dominated by the vector meson poles leading to a characteristic behavior 
also in the space-like momentum region.

When assessing the influence of the different momentum dependent dressing functions
of the DSE approach as compared to ENJL we found partial cancellations. Dressing effects
due to the non-trivial wave-function on the level of $50\%$ are
cancelled by opposite effects due to the dressing of the gauge part of the vertex.
An important effect that is not cancelled is the one of the dynamical mass function.
We found that this function is not tested predominantly at its large infrared
plateau of $M(0)\sim0.48$~GeV but rather at smaller values at intermediate momenta
commensurate with $M~\sim0.2$~GeV. This gives rise to a larger contribution than
expected from constituent quark-loop calculations. Our finding may serve to explain 
the surprisingly small (constituent) quark masses needed in chiral models to obtain 
sensible results \cite{Greynat:2012ww}. Our most interesting result, however,
is related to the transverse part of the vertex. These are of high interest
because they dynamically include the phenomenology of vector meson dominance and are 
expected to reduce the overall light-by-light contribution. As in the ENJL model
the corresponding behavior of the transverse part of the vertex in the DSE-approach is 
generated dynamically. A key difference is, however, that in the DSE case the dependence 
on the relative momentum of the quarks is taken into account. On the level of mesons 
(as e.g. the $\rho$-meson that is the vital ingredient in VMD) this takes the distribution 
of quarks inside the bound state into account, which is not the case in the ENJL 
model \cite{Tandy:1997qf}. The inclusion of this effect,
either in the full DSE calculation or by a suitable modification of the VMD-term in 
the ENJL model has important consequences: the reduction of the quark-loop contribution 
to $a_\mu$ due to VMD effects, observed in previous calculations \cite{Bijnens:1995xf}, 
is drastically reduced. 

Thus we can pinpoint the differences between the DSE and ENJL calculations to be down
to the contact interaction limiting the momentum dependence of dressing functions. We
believe that ignoring the relative quark momentum in the quark-photon vertex overestimates
the suppression that the transverse part of the vertex provides and thus lowers
significantly it's numerical contribution to the anomalous magnetic moment of the muon.

Our present best result for the quark-loop contribution has been discussed in section
\ref{best}. Combined with our result for the pseudoscalar meson exchange diagram from
Ref.~\cite{Goecke:2010if}, $a_\mu^{LBL;PS} \approx (81 \pm 2) \times 10^{-11}$, 
we arrive at the estimate 
\begin{align}
  a_\mu^{\mathrm{LBL}} = (188 \pm 4) \times 10^{-11},
  \label{final}
\end{align}
for the total LBL contribution. Again, the error is purely statistical due to our numerics.
Since at present any guess of the systematic error of this number is clearly subjective 
(due to the omission of terms in the quark-photon vertex), we do not attempt such an estimate. 

It has to be emphasized, however, that our determination of the quark loop contribution 
to light-by-light, Eq.~(\ref{final}) is by far not complete since several terms in the vertex dressing are 
still missing. The study if the influence of these terms is an important task for the future.
Nevertheless we hope that the systematics of the present work serves to give the reader 
a better understanding of the technical and physical mechanisms at work in the complicated 
light-by-light scattering contribution. Furthermore we have shown that our best results
at present are stable under variations of the photon gauge parameter and therefore serve
as an important intermediary step towards a full calculation to come. 

\vspace*{3mm}
{\bf Acknowledgments}\\
We are grateful to Fred Jegerlehner and Andreas Nyffeler for discussions.
This work was supported by the Helmholtz International Center for FAIR within 
the LOEWE program of the State of Hesse, the Helmholtz Young Investigator Group 
under contract VH-NG-332 and DFG under contract FI 970/8-1, and the Austrian
Science Fund FWF under project M1333-N16.

\appendix
\section{Quark-photon vertex}
In this appendix we give the 
explicit basis for the quark-photon vertex.
It was taken from Ref. \cite{Maris:1999bh}.
\begin{align}
  \begin{split}
  L_\mu^1 &= \gamma_\mu								\\
  L_\mu^2 &= 2k_\mu\sh{k}							\\
  L_\mu^3 &= 2\i k_\mu								\\
  L_\mu^4 &= \i [\gamma_\mu,\gamma_\nu]k_\nu					\\
  T_\mu^1 &= \gamma_\mu^T							\\
  T_\mu^2 &= [k_\mu^T \sh{k}^T-\frac{1}{3}\gamma_\mu^T (k^T)^2]/k^2		\\
  T_\mu^3 &= k_\mu^T \Sh{P} P\cdot k /(k^2P^2)					\\
  T_\mu^4 &= -(\gamma_\mu^T[\Sh{P},\sh{k}^T]+2k_\mu^T\Sh{P})/2 k		\\
  T_\mu^5 &= \i k_\mu^T/k							\\
  T_\mu^{6} &= \i[\gamma_\mu^T,\sh{k}^T]P\cdot k /k^2				\\
  T_\mu^{7} &= \i[\gamma_\mu^T,\Sh{P}]\left(1-\frac{(P\cdot k)^2}{P^2 k^2}\right)-
  		2 T_\mu^{8}							\\
T_\mu^{8} &= \i k_\mu^T\sh{k}^T\Sh{P}/k^2.
\end{split}
  \label{eqn:QPVMarisTandyBasis}
\end{align}
Here $P$ is the total momentum (photon momentum) and
the relative momentum is $k=(k_++k_-)/2$ were
$k_\pm$ are the quark momenta. The symbol $T$ denotes
transversality with respect to $P$. Furthermore,
$k=\sqrt{k^2}$. The twelve dressing functions corresponding
to the tensor structure (\ref{eqn:QPVMarisTandyBasis})
are denoted as $\lambda^{(i)}$, $(i=1,2,3,4)$ corresponding
to the $L^{(i)}_\mu$ and $\tau^{(j)}$ with $(j=1,\ldots,8)$
corresponding to the $T^{(j)}_\mu$, see Eq. (\ref{eqn:VertexDecomposition}).
The dressings $\lambda^{(i)}$ are determined by the
WTI
	\begin{align}
	  \i P_\mu \Gamma_\mu(P,k) &= S^{-1}(k_-) - S^{-1}(k_+).
	  \label{eqn:QEDWTI}
	\end{align}
and regularity demands \cite{Ball:1980ay}.
The resulting vertex construction
is referred to as the Ball-Chiu construction
\begin{align}
  \Gamma_\mu^{\mathrm{BC}}(P,k)&=
   \gamma_\mu \Sigma_A + 2\sh{k} k_\mu \Delta_A
   +\i k_\mu \Delta_B,
  \label{eqn:BCVertex}
\end{align}
where the symbols 
\begin{xalignat}{2}
  \Sigma_F&=\frac{F(k_+^2)+F(k_-^2)}{2}
  &\Delta_F&=\frac{F(k_+^2)-F(k_-^2)}{k_+^2-k_-^2},
\end{xalignat}
have been used and $A$ and $B$ are the quark dressings.
Note that $\lambda^{(4)}$ is identically zero.

\section{Derivation of the hadronic four-point function}\label{appB}
\begin{figure}
  \begin{center}
    \includegraphics[width=0.5\textwidth]{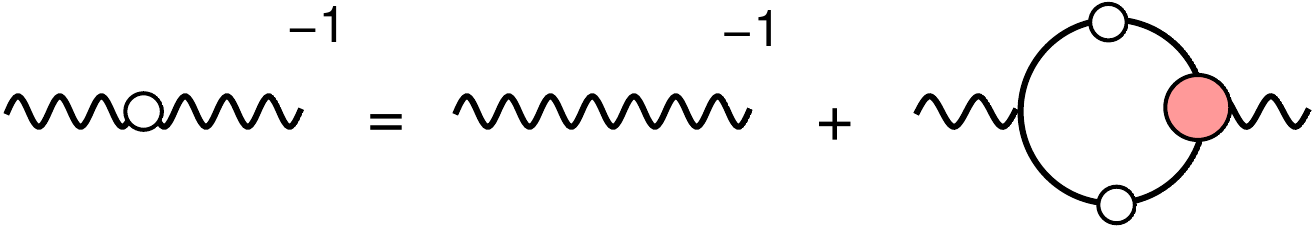}
  \end{center}
  \caption{The photon DSE.
  	Full Propagators are denoted by white blobs and the red blob marks the
	full 1PI fermion-photon vertex.}
  \label{fig:photonDSEAppendix}
\end{figure}

In the present appendix we present the derivation of the
photon four-point function that is needed in the light-by-light
scattering contribution to the muon $g-2$. The four-point function
is obtained by taking two derivatives of the full inverse photon
propagator
\begin{align}
  \begin{split}
  \Pi_{\mu\nu\alpha\beta}&= \frac{\delta^4 \Gamma[A]}{\delta A_\mu \delta A_\nu \delta A_\alpha \delta A_\beta} \\
  & = \frac{\delta^2}{\delta A_\mu \delta A_\nu}\frac{\delta^2 \Gamma[A]}{\delta A_\alpha\delta A_\beta}
  = \frac{\delta^2}{\delta A_\mu \delta A_\nu}\left( D_{\alpha \beta} \right)^{-1}.
\end{split}
  \label{eqn:DefOf4ptFunctionInTermsOfPhoton2ptFunction}
\end{align}
In order to proceed we consider the photon DSE
that is presented in Fig. \ref{fig:photonDSEAppendix}.
We now take two further derivatives of the photon DSE. By the virtue
of the approximation we consider, namely rainbow-ladder truncation of QCD,
the photon is not a dynamical field in our formalism. The photon self-energy
that we calculate is performed using quark and quark-photon vertex equations
that do not take into account the photon as a dynamical object, but rather as an external
background field. Internally all objects just include dynamical quarks and model gluons.
This is why the application of further derivatives on the photon DSE is consistent
within our approximation. Furthermore this operation leaves the internal consistency
of the truncation intact. In particular the consistency with chiral symmetry
and electro-magnetic current-conservation in form of WTI's is not destroyed.
This technique is also referred to as 'gauging' \cite{Haberzettl:1997jg}.
For a detailed description of this procedure in particular in
rainbow-ladder truncation see \cite{Eichmann:2011ec}.\\

An object that will be needed on several occasions is the quark anti-quark photon-photon
vertex. Within rainbow-ladder approximation this object can be exactly written
in terms of quark-photon vertices and the T-matrix \cite{Eichmann:2011ec} 
\begin{align}
  \parbox{0.3\textwidth}{
		\includegraphics[width=0.3\textwidth]{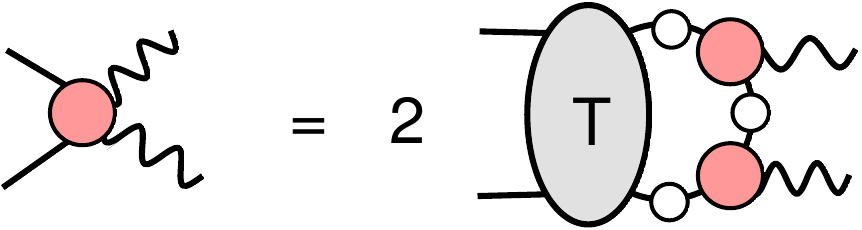},
  }
  \label{eqn:DefOf2quark2photonVertex}
\end{align}
where the factor of two is not a symmetry factor but rather a convenient
way to denote that the two possible photon permutations are included.
Application of one derivative to the photon DSE in Fig. \ref{fig:photonDSEAppendix}
gives
\begin{align*}
  \parbox{0.45\textwidth}{
	\includegraphics[width=0.45\textwidth]{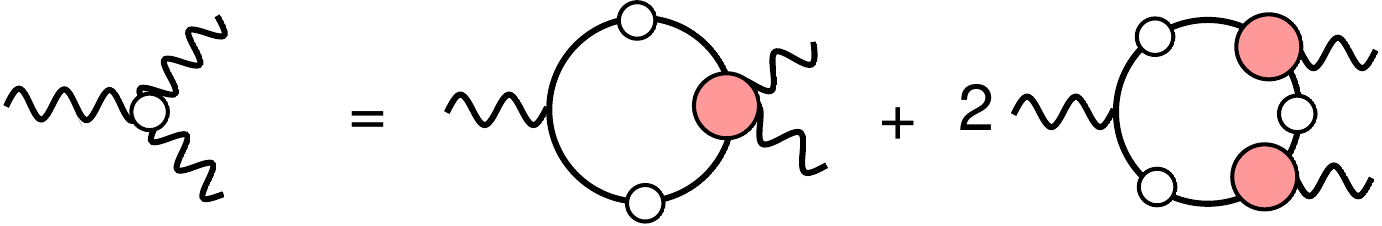}.
  }
\end{align*}
The bare inverse photon does not contribute. The derivative has to be applied
to the quark-photon vertex and the two dressed quarks of the photon self energy.
Since there are two quarks in the loop the second diagram comes again with two
permutations that correspond to one diagram with straight photon legs
and one with crossed ones. This is signaled by the factor $2$. Now the application 
of the relation shown in Eq. (\ref{eqn:DefOf2quark2photonVertex}) results in
\begin{align*}
  \parbox{0.45\textwidth}{
	\includegraphics[width=0.45\textwidth]{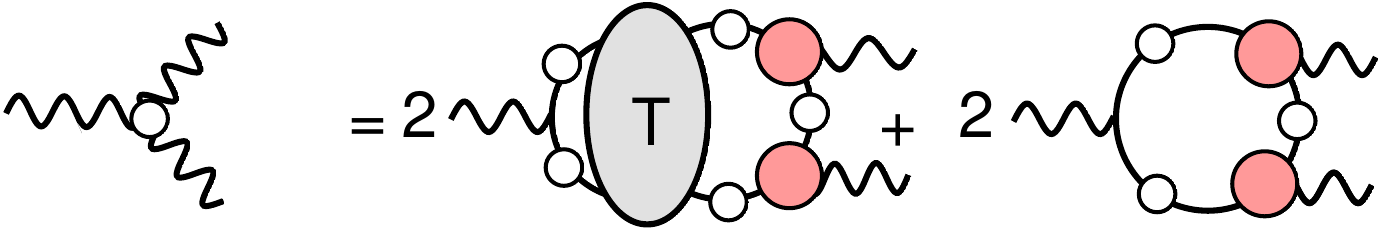}.
  }
\end{align*}
Next we take into account the relation between the T-matrix and the G-matrix 
\begin{align}
  G = SS + SSTSS,
  \label{eqn:GandTMatrix}
\end{align}
which says nothing but that the T-matrix is the amputated connected part
of the G-matrix. In Eq. (\ref{eqn:GandTMatrix}), $S$ is the quark propagator
and multiplication indicates contraction of dirac-indices and integration
over momentum arguments.
The inhomogeneous 
part that involves just two propagators ($SS$ in Eq. (\ref{eqn:GandTMatrix}))
cancels the second diagram on 
the right side of the equation above and we are left with
\begin{align*}
  \parbox{0.3\textwidth}{
	\includegraphics[width=0.3\textwidth]{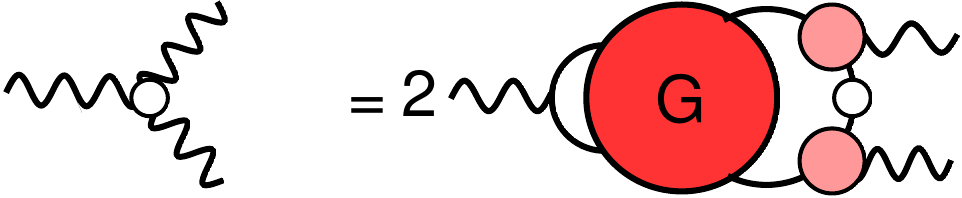}.
  }
\end{align*}
Now the final step is to remember that the bare quark-photon vertex
together with the G-matrix corresponds to a dressed quark-photon vertex
and to external quark-leg dressings. Finally we obtain the 
consistent representation of the hadronic part
of the three-photon vertex
\begin{align*}
  \parbox{0.3\textwidth}{
	\includegraphics[width=0.3\textwidth]{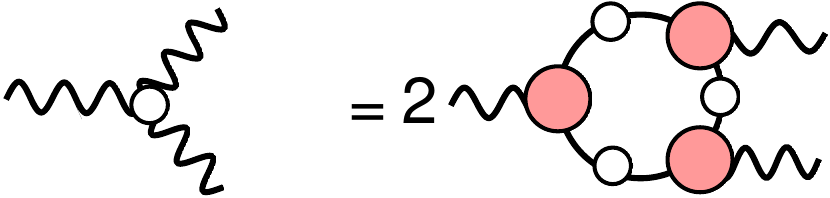},
  }
\end{align*}
which is not zero as long as the external photon background-field
is present. The factor $2$ can also be interpreted as representing
the two possible orientations of the quark loop.\\

Now the four-point function is obtained by taking a further derivative
with respect to the external photon field. Again the derivative can
act on quark-photon vertices and the quark propagators. The result is
\newpage
\begin{align*}
  \parbox{0.45\textwidth}{
	\includegraphics[width=0.45\textwidth]{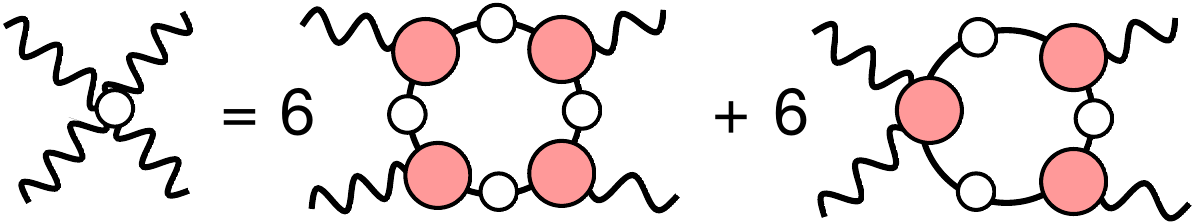},
  }
\end{align*}
where the additional factor three is caused by the three possible vertices
and three propagators respectively. Using again relation
(\ref{eqn:DefOf2quark2photonVertex}) we arrive at
\begin{align*}
  \parbox{0.45\textwidth}{
	\includegraphics[width=0.45\textwidth]{4photonvertex2}.
  }
\end{align*}
Up to this point no approximations on top of rainbow-ladder
truncation of QCD have been made. The last two representations
of the four-point function are thus truly consistent with the rainbow-ladder
photon self-energy shown in Fig. \ref{fig:photonDSEAppendix}.\\

\end{document}